\documentclass{aa}

\usepackage{natbib}

\usepackage{graphicx}
\usepackage{txfonts}
\usepackage{enumerate}
\usepackage[usenames]{color}

\newcommand{\psfprime}{{\hbox{PSF}^\prime}}

\begin{document}

\title{The power spectrum of solar convection flows from
  high-resolution observations and 3D simulations}

\titlerunning{Power spectra of solar convection flows}
\authorrunning{L. Yelles Chaouche \& F. Moreno-Insertis \& J. A. Bonet}

\author{L. Yelles Chaouche\inst{\ref{inst1}, \ref{inst2}} \and
  F. Moreno-Insertis \inst{\ref{inst1}, \ref{inst2}} 
\and J. A. Bonet\inst{\ref{inst1}, \ref{inst2}}}

\institute{Instituto de Astrofisica de Canarias, Via Lactea, s/n,
  38205  La Laguna (Tenerife), Spain\label{inst1}
  \and
Dept.~of Astrophysics, Universidad de La
  Laguna, 38200 La Laguna (Tenerife), Spain\label{inst2}}

\date{Received:  / Accepted: }

\abstract
{Understanding solar surface magnetoconvection requires the study of the
  Fourier spectra of the velocity fields. Nowadays, observations
  are available that resolve very small spatial scales, well into the
  subgranular range, almost reaching the scales routinely resolved in
  numerical magnetoconvection simulations.  Comparison of numerical and
  observational data at present can provide an assessment of the validity of
  the observational proxies.}
{Our aims are: (1) to obtain Fourier spectra for the photospheric velocity fields using the
  spectropolarimetric observations with the highest spatial resolution so far
  ($\sim 120$ km), thus reaching for the first time spatial scales well into
  the subgranular range; (2) to calculate corresponding Fourier spectra
  from realistic 3D numerical simulations of magnetoconvection and carry out
  a {{\it proper comparison}} with their observational counterparts
  considering the residual instrumental degradation in the observational
  data; and (3) to test the observational proxies on the basis of the
  numerical data alone, by comparing the actual velocity field in the
  simulations with synthetic observations obtained from the numerical boxes.}
{(a) For the observations, data from the SUNRISE/IMaX spectropolarimeter are
  used. (b) For the simulations, we use four series of runs obtained with the
  STAGGER code for different average signed vertical magnetic field values
  ($0$, $50$, $100$, and $200$ G). Spectral line profiles are synthesized from
  the numerical boxes for the same line observed by IMaX (Fe I 5250.2 \AA)
  and degraded to match the performance of the IMaX instrument. Proxies for
  the velocity field are obtained via Dopplergrams (vertical component) and
  LCT (local correlation tracking, for the horizontal component). Fourier
  power spectra are calculated and a comparison between the synthetic and
  observational data sets carried out. (c) For the internal comparison of the
  numerical data, velocity values on constant optical depth surfaces are used
  instead of on horizontal planes.}
{A very good match between observational and simulated Fourier power spectra
  is obtained for the vertical velocity data for scales between $200$ km and
  $6$ Mm. Instead, a clear vertical shift is obtained when the synthetic
  observations are not degraded to emulate the degradation in the IMaX
  data. The match for the horizontal velocity data is much less impressive
  because of the inaccuracies of the LCT procedure.
\ Concerning the internal comparison of the direct velocity values of the
numerical boxes with those from the synthetic observations, a high
correlation ($0.96$) is obtained for the vertical component when using 
the velocity values on the $\log \tau_{500} = -1$ surface in the box. The
corresponding Fourier spectra are near each other. 
A lower maximum correlation ($0.5$) is reached (at $\log \tau_{500} = 0$) for the
horizontal velocities as a result of the coarseness of the LCT procedure.
Correspondingly, the Fourier spectra for the LCT-determined velocities 
is well below that from the actual velocity components.}
{As measured by the Fourier spectra, realistic numerical simulations of
  surface magnetoconvection provide a very good match to the observational
  proxies for the photospheric velocity fields at least on scales from
  several Mm down to around $200$ km. Taking into account the spatial and
  spectral instrumental blurring is essential for the comparison between
  simulations and observations. Dopplergrams are an excellent proxy for the
  vertical velocities on constant-$\tau$ isosurfaces, while LCT is a much
  less reliable method of determining the horizontal velocities.}

\keywords{Sun: Photosphere --- Sun: Granulation --- Convection ---
  Magnetohydrodynamics --- Turbulence}

\maketitle

\section{Introduction}\label{sec:introduction}

The convective flows in the outermost layers of the solar interior and
photosphere play an essential role in the dynamics and magnetism of the solar
envelope and atmosphere. Their kinetic energy is distributed over a very wide
range of space- and timescales \citep[see, e.g.,][]{nordlund2009}, as results
from (a) the strong stratification in the topmost tens of Megameters below
the surface, leading to the appearance of  convection cells from granular
  to supergranular size; and (b) the very high Reynolds numbers of those
flows, which lead to turbulence and cascading of energy from the
  granular size downward to small scales before reaching the dissipation
  range.  Solar convection provides a prime example of the need for
interaction between observations and numerical simulations.  The most recent
advances in observational facilities now permit the observation of
structures, flows, and fields with very high spatial resolution,  down to
  $\approx 100$ km.
On the other hand, the numerical experiments and simulations of the past ten
years, using the power of recent supercomputing installations, have reached
high realism concerning the physical processes studied in the solar interior
and photosphere (including material properties and detailed radiative
transfer) with numerical grid spacing down to the order of $10$ km
\citep[e.g.,][]{2006ApJ...642.1246S, 2007A&A...465L..43V,
  2008ApJ...684L..51J}. In addition, for the large scales, realistic
simulations are now being produced on boxes with sides spanning several tens
of Megameters in the horizontal directions, hence including a few
supergranular cells in them \citep[e.g.,][]{2011SoPh..268..271S}. The
simulations of convection with different numerical codes seem to agree
reasonably well \citep{Beeck_etal_2012}.

A crucial tool to understanding the nature of the photospheric flows is their
Fourier power spectrum.  Application of this tool has been made along the
years to both observational and numerical simulation data, mostly separately,
but sometimes also comparing results from the two sources
\citep[e.g.,][]{Stein_Nordlund1998, 2006ASPC..354..109G, Georgobiani2007, 
  Kitiashvili_etal_2012}. For the observations, proxies have to be used for
the velocity components, like Doppler shifts of spectral lines for the
vertical velocity \citep[e.g.,][]{Hathaway_etal_2000, Stein_etal_2006,
  Georgobiani2007, rieutord2010, Kitiashvili_etal_2012, Katsukawa2012} or
feature displacement on series of 2D continuum or Doppler maps
\citep[e.g.,][]{rieutord2001, rieutord2010, Stein_etal_2006,
  2006ASPC..354..109G, Georgobiani2007, 2012A&A...540A..88R, Goode_etal_2010,
  Abramenko2012, Kitiashvili_etal_2012}. For the simulations, the velocity
values on horizontal cuts in the numerical box have been Fourier-analyzed
\citep[][]{Stein_Nordlund1998, rieutord2001, Stein_etal_2006,
  2006ASPC..354..109G, Georgobiani2007, Kitiashvili_etal_2012}.  In the
current paper we would like to push the limits of the comparative analysis
between simulations and observations in  three directions:

\begin{enumerate}[(i)]

\item \label{enum:i} by using Fourier spectra from observational proxies for
  the photospheric velocity field (Doppler, Local Correlation Tracking)
  using as a basis recent observations with very high spatial resolution;

\item \label{enum:ii} by comparing them with Fourier spectra from numerical
  simulations but taking great care that the comparison is between homogeneous
  datasets, i.e., by first obtaining {\it synthetic observations} from the
  simulations and then modifying them to reproduce the instrumental
  degradation suffered by the real observations; and

\item \label{enum:iii} by checking the closeness of the synthetic
  observational proxies to the actual velocity vector components in the
  numerical box when using surfaces of constant optical depth
  ($\tau_{500}$=const) that correspond to the formation region of the
  spectral lines.

\end{enumerate}

For the first aspect, observations with spatial resolution as high as
one-tenth of the average granular size are available at present.  Using
Hinode/SOT data (spatial resolution: $200$ km), \cite{rieutord2010} have
explored the power spectra of the vertical velocity down to scales of $400$
km (and of the horizontal velocity in a range from $2.5$ Mm to $76.3$ Mm),
but no comparison to numerical simulations was provided.
\citet{Goode_etal_2010} obtained continuum images providing information down
to very small scales of about $90$ km using speckle-reconstruction techniques
with the NST/BBSO telescope. From them, they could obtain horizontal-velocity
maps using Local Correlation Tracking (LCT) techniques. In the present paper
we take advantage of the availability of time series of spectropolarimetric
observations with unprecedented spatial resolution (near $120$ km) obtained
using the IMaX instrument aboard the SUNRISE balloon mission
\citep{martinezpillet_etal_2011, solanki_etal_2010}.
These data allowed the determination of the line-of-sight velocity,
continuum maps (used to compute the horizontal velocity by applying LCT) and of
the full magnetic field vector with a field of view of $32 \times 32$ Mm$^2$
and for periods of half an hour with a $33.3$-sec cadence.

For the second aspect, comparisons so far have
been done almost exclusively by using observational proxies, on the one hand,
and 2D maps on horizontal planes in the numerical box, on the other. 
Yet, as done in the present paper, using spectral synthesis it is possible to
obtain synthetic observations from the simulations that can then be subjected
to the same sort of quality degradation that the solar spectrum suffers when
going through the observational equipment: for that, one only needs to carry
out a convolution of the synthetic spectra with a Point Spread Function (PSF)
in the spatial domain and with the Monochromatic Spectral Response Function
(MSRF) in the spectral domain adequate to the specific equipment used to
obtain the observations. In this way, the comparison can be carried out using
the same kind of proxies on either side.

Finally, for the third aspect, the numerical datasets and techniques
used in this paper allow us to approach the question of the closeness of the
observational proxies (Dopplergrams and horizontal-velocity maps from feature
tracking) to the actual velocity fields.  So, it is of interest to check if
the observational proxies are near the actual velocity field when both are
obtained from the numerical results.  The actual velocity field data have to
be taken from a representative surface in the box, for which we choose an
isosurface of the optical depth at $500$ nm, $\tau_{500}$,  which seems more
  appropriate than a horizontal plane in the numerical box.  This is because,
  following basic spectral line formation theory \citep[e.g.,][]{gray2005},
  the lines used for the Doppler velocity determination here are expected to
  be formed roughly in the same $\tau_{500}$--range for all vertical columns
  in the numerical box or in the observations (see further explanations in 
  Sect.~\ref{sec:velocities_at_grid_points}). 

The layout of the paper is as follows: Sect.~\ref{sec:observations} and
\ref{sec:simulations} describe the observational and numerical datasets,
respectively, used in the paper. Section~\ref{sec:IMaXconditions} details the
procedure whereby we gain synthetic observations from the numerical boxes and
the instrumental degradation that we have to apply to make these synthetic
observations coherent with the observational data. The core of the paper are
Sect.~\ref{sec:vertical} through \ref{sec:velocities_at_grid_points}.
Sect.~\ref{sec:vertical} and \ref{sec:horizontal} discuss the Fourier power
spectra obtained for the vertical and horizontal velocity distributions,
respectively. Section~\ref{sec:velocities_at_grid_points} compares the
observational proxies with the numerical data taken on $\tau_{500}$
isosurfaces. Sections~\ref{sec:summary} and \ref{sec:conclusions} provide a discussion and
conclusions.

\section{Observations}\label{sec:observations}

\begin{figure}[!ht]
\begin{center}
\includegraphics[width=0.5\textwidth]{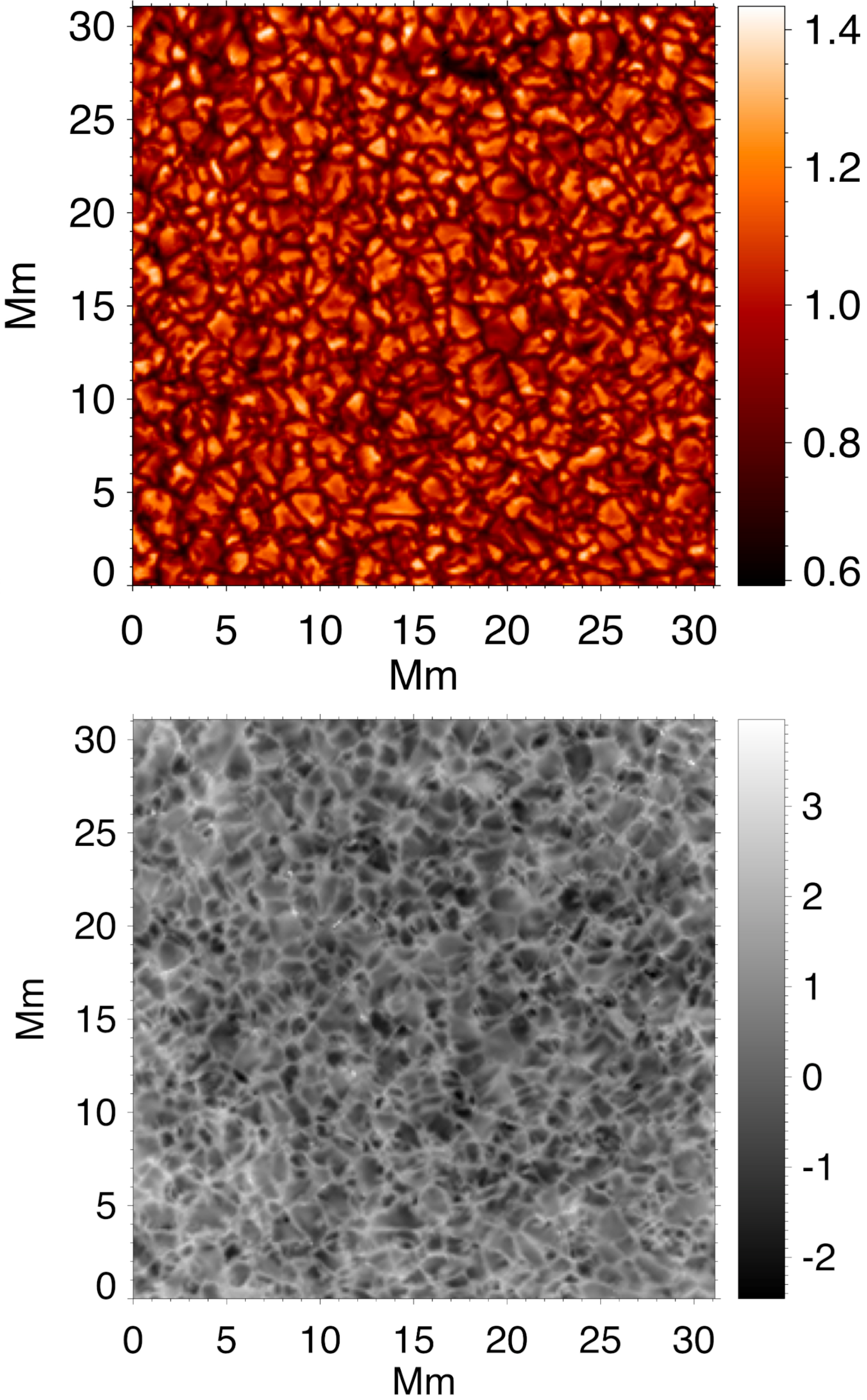}
\end{center}
\caption{Top: Continuum intensity observed by the SUNRISE/IMaX instrument
  near the spectral line Fe I 5250.2 \AA\ .  Bottom: Doppler (vertical)
  velocity computed using the Stokes-I signal of the Fe I 5250.2 \AA\ line.
  The velocity is expressed in km/s. Details about granules and inter
  granular lanes can be well appreciated.} \label{fig1}
\end{figure}

For this study we use sequences of images recorded with IMaX near the solar
disk center on 2009 June 9. Images with pixel size $39.9$ km were taken at
five wavelengths along the profile of the magnetically sensitive FeI $5250.2$
\AA\ line, located at $\pm 80$ m\AA, $\pm 40$ m\AA\ from line center, and
continuum at $+227$ m\AA. The images were reconstructed considering the
instrumental aberrations calibrated by the phase diversity method. The
estimated circular polarization noise is $5\times 10^{-4}$ in units of the
continuum wavelength for non-reconstructed data and three times larger for the
reconstructed one.  IMaX has a spectral resolution of $85$ m\AA\ and a
spatial resolution of $110$ -- $130$ km \citep[see][where further details
  about data acquisition and reduction procedures are
  given]{martinezpillet_etal_2011}.  We use two time series with a field of
view (FOV) of $32 \times 32$ Mm$^2$ and cadence $33.3$ sec; the first series
comprises $42$ snapshots ($23$ minutes) and the second one $58$ snapshots
($32$ minutes).  The top panel of Figure~\ref{fig1} displays a map of the
normalized continuum intensity of one of the images. This map indicates that
the contrast is particularly high thanks to the high resolution and the
virtually seeing-free conditions of the IMaX instrument.  The bottom panel of
Figure~\ref{fig1} shows the corresponding Doppler (vertical) velocity map. It
shows downflows (positive values in the greyscale sidebar) associated with
narrow inter-granular lanes, and upflows associated with the body of
granules.

\section{Simulations and spectral synthesis}\label{sec:simulations}

For the simulations we use series of (magneto)convection runs calculated with
the Stagger code, developed by A. Nordlund (see a description of the code in
the paper by \citealt{Beeck_etal_2012}). This code includes a realistic
equation of state with partial ionization effects; the radiation transfer
problem is solved in the code using the frequency-binning method.  In the
particular runs in this paper, four frequency bins were used, and the
transfer equation is solved along nine rays through each grid point.

The integration box has a size of $6 \times 6$ Mm$^2$ in the horizontal
directions and $2.5$ Mm in the vertical direction. The photosphere, or,
  more precisely, the average $\tau_{500}=1$ level, is located some $430$ km
  below the top lid of the box, with small variations of $\pm 10$ km for the
  different time series described below. The numerical grid has $252 \times
252$ (horizontal) and $126$ (vertical) points. The grid point separation in
the horizontal directions is uniform and equal to $23.8$ km, with periodic
conditions on the side boundaries.  A non-uniform grid is used in the
vertical direction, with vertical grid spacing of $15$ km near the
photosphere.  The dataset used in this paper consists of four time series
each for a different value of the average signed vertical magnetic field 
strength $\langle B_z \rangle$ (with the averages taken on single horizontal
cuts), namely $0$ G (HD case) and $50$ G, $100$ G and $200$ G
(magnetoconvection cases). The convection runs were initially started with a
uniform vertical magnetic field of the nominal strength for each
series. Given the conservation of magnetic flux and the periodicity condition
on the side boundaries, the average $\langle B_z \rangle$ is uniform across
all heights and constant in time.  The convective motions are pursued over a
timespan of $30$ solar minutes in each series; for the present paper we used
snapshots stored with a $30$-sec cadence. In all cases, the $30$ solar-min
series used in the paper are extracted from the convection runs once a
statistically-stationary state is obtained.  These convection runs have also
been used for the solar abundance studies of \citet{Fabbian_etal_2010,
  Fabbian_etal_2012} and in the paper by \citet{Beck_etal_2012}, where
further details about the numerical model can be found.

A central role in this paper is played by the synthetic observations obtained
from the numerical runs. To calculate them, we synthesize the electromagnetic
spectrum for a few spectral lines emitted from the plasma in the individual
columns of the numerical boxes.  For the spectral synthesis we used the
Nicole code \citep{NicoleSocas}. For each vertical column in the box and for
every snapshot in the series the outgoing spectrum is calculated for the
\ion{Fe}{1} 5250.20 \AA\ line (the IMaX line) and for the Fe I 5576.09
\AA\ line.  To feed the Nicole code, a segment of each column stretching
between $\log(\tau_{500}) = -4 $ and $\log(\tau_{500})= 2 $ is selected and
finely re-gridded in equal intervals of $\log\tau$. This allows an optimum
opacity resolution and yields smooth continuum intensity maps in each box, a
necessary condition for proper application of the LCT method (see
Sect.~\ref{sec:IMaXconditions}). The columns are assumed to coincide with the
LOS, i.e., the boxes are assumed to be placed at disc center. The lines are
synthesized assuming LTE and using the following atomic parameters: for the
$5250.20$ \AA\ line we took excitation potential $\chi_l = 0.121$ eV and log(gf)
$= -4.94$; for the $5576.09$ \AA\ line we used $\chi_l = 3.43$ eV and log(gf)
$=-0.94$. In either case we used the Uns\"old collisional line broadening
formalism with no enhancement. For a discussion of the synthetic spectra
obtained from these snapshots and their closeness to the solar spectrum
see the recent publications by \citet{Fabbian_etal_2010, Fabbian_etal_2012}
and \citet{Beck_etal_2012}.

\section{Adjusting simulations to IMaX conditions}\label{sec:IMaXconditions}

In order to compare the synthetic observations gained from the numerical
simulations with the real observations from IMaX it is necessary to modify
the former to take into account the residual instrumental degradation in the
original data available from the SUNRISE mission. Therefore we have to apply
the following steps to the simulated data:

(i) The original IMaX data had been corrected for
  instrumental optical aberrations by deconvolving with a PSF derived from
  phase diversity calibration thus achieving an effective resolution limit of
  $0.15$-$0.18$ arcsec as reported by \citealt{martinezpillet_etal_2011}.  However,
  this calibration does not model the effect of the stray light (far wings of
  the PSF). Thus, to render the synthetic observations comparable with the
  original IMaX data, the Stokes-$I$ spectra coming out of the numerical box
  are convolved spatially with an additional point spread function, called
  $\psfprime$ in the following, to simulate the stray light contamination due
  to photons scattered by the various optical components of the instrument
  \citep{Wedemeyer2008, Danilovic_etal_2008, Wedemeyer_Vandervoort_2009, 
    Scharmer_etal_2010, 
    Beck_etal_2012}. The mathematical expression we use to model this effect
  is a combination of a pulse (i.e., a Dirac-delta) and a Lorentzian function
  ${\cal L}{(A,r)}$ with relative weight $w$, 
\begin{equation}\label{eq:PSF}
\psfprime (r) = w\, \delta (r) + (1-w) \, {\cal L}(\alpha,r) \;,
\end{equation}
\noindent where $\alpha$ is the parameter ruling the width of the Lorentzian.
${\cal L}(\alpha,r)$ is chosen so that the area integral of $\psfprime$ is
unity.  The values of $w$ and $\alpha$ are tuned so that the contrast of the
synthetic continuum images matches the observed contrast of the IMaX time
series.  Because of the presence of a Dirac-delta in the definition of
$\psfprime$, the cutoff wavenumber of its Fourier transform (the optical
transfer function) is at infinity.  Consequently, the validity of the
comparison between theoretical and observational power spectra must be
restricted to the range up to the effective cutoff achieved by IMaX, i.e., to
spatial scales of $110$--$130$ km.

(ii) The synthetic spectra are convolved in the spectral dimension with a
  monochromatic response function (MSRF). A Gaussian profile is used here. It has a
  FWHM $= 85$ m\AA\ in order to take into account the effect of the filter
  used to measure the spectra.  For the sake of
    conciseness we will refer to the degradations described so far as
    {\it the residual instrumental degradation.}

(iii) The
resulting spectra are resampled spectrally at five wavelength positions
corresponding to the location of the IMaX spectral sampling (see
Sect.~\ref{sec:observations}).  

(iv) The vertical velocity is determined by computing the Doppler shift of
  the spectral lines.  To this end, a Gaussian fit is performed on the
  Stokes-I profiles, and the location of the minimum of the Gaussian curves
  with respect to the line center at rest is taken as the Doppler shift,
  which is then converted into a Doppler velocity. This procedure is used
  both for observations and simulations to make them comparable. In the case
  of the simulations, it is possible to compare the velocity determined by
  this method with the one obtained from the center of gravity of the full
  profile (before spectral resampling) with very high spectral resolution
  ($7.5$ m\AA). From the comparison we conclude that the two methods lead
  to quite similar velocities.

(v) The horizontal velocity is computed using Local Correlation Tracking
  (LCT) applied to successive continuum images. The LCT routine
  \citep{welsch_etal_2004} is performed in local windows weighted by a
  Gaussian function with FWHM=$320$ km and yields the two components of the
  horizontal velocity.  Since no spectra are used in this case, to bring the
  simulated continuum images to observational conditions, we need only
  perform the step (i) described above. Specifically, we compute the
continuum near the spectral line Fe I 5250.2 \AA\ (the IMaX line) and
convolve the resulting images with the $\psfprime$ function introduced in
(i).

(vi) To complete the adjusting procedures between observations and
simulations and to make the results comparable with other
observations and simulations we use a p-mode filtering for both the
vertical and the horizontal velocity components.  Following the
procedure indicated by \citet{title1989}, the p-modes are removed by applying
a subsonic filter which removes modes located in the portion where
$\omega/k > V_{ph}$ in the ($k,\omega$) diagram, with $V_{ph}$ being a chosen
cutoff velocity. Here we use $V_{ph} =4$ km/s for most of the figures.

\section{Fourier power spectrum and velocity spectrum:
  definitions}\label{sec:fourier_definitions} 

For a given component of the velocity, $v_i$, in a given map, its power
spectrum $P_i(k)$ is defined via Parseval's theorem as
\begin{equation}\label{eq:powerspectrum}
P_i(k) = k\int_0^{2\pi} 
\frac{\left| \strut [{\cal F}(v_i)](\vec{k})\right|^2}{A}
\;d\theta\;,
\end{equation}
with ${\cal F}(v_i)$ the two-dimensional Fourier transform of $v_i$,
$A$ the area of the spatial
domain, and $k$ and $\theta$ the module and orientation angle, respectively,
of the wavevector $\vec{k}$. When speaking of the power spectrum of the
horizontal velocity below, we mean the sum of the spectra for the two
horizontal velocity components. To illustrate the physical meaning of the
spectrum (\ref{eq:powerspectrum}), we recall that in a situation with
uniform density the kinetic energy spectrum of the map, $E(k)$, is given by
\begin{equation}\label{eq:kinetic_energy_spectrum}
E(k) = \frac{1}{2} \left[\strut P_x(k) + P_y(k) + P_z(k)\right]\;,
\end{equation}
so the total kinetic energy per unit mass and area in the given map would be
given by $\int_0^{\infty} E(k)\, dk$.  In fact, in the following, we will
mostly use the so-called {\it velocity spectrum} which is generically defined
as 
\begin{equation}\label{eq:velocityspectrum}
V(k) = [k\, P(k)]^{1/2}\;,
\end{equation}
where, depending on the context, for $P(k)$ the spectrum of a given velocity
component or of the horizontal velocity  will be used. Trivially, the
velocity spectra $V(k)$ have the
dimension of a velocity  
and fulfill
\begin{equation}\label{eq:velocityspectrum_and_kineticenergy}
  \int_0^{\infty} E(k)\,dk = \frac{1}{2}\,\int_0^{\infty} \left[\strut V_x^2(k) +
    V_y^2(k) + V_z^2(k)\right]\;d\ln k \;,
\end{equation}
so they provide a measure for the velocity amplitude corresponding to a given
spatial scale 
\citep[see also][]{nordlund2009}.  To characterize length scales we will use
the wavelength of the Fourier mode $\lambda = 2\pi/ k $. In the figures we
will also use the spherical harmonic wavenumber, $m =2 \pi R_{\sun} /
\lambda$, with $R_{\sun}$ the solar radius. Following
  our use of the Fast Fourier Transform algorithm on finite-domain data, we
  will be using summations over a finite number of terms to calculate the
  spectra of Eq.~\ref{eq:powerspectrum} or the total kinetic energy in the
  box. 

When calculating the Fourier spectra we compensate for the lack of
periodicity of the real observations by zero-padding the outer rows and
columns of the velocity map. To avoid the spurious high-frequency
contributions caused by the strict application of a rectangle function with
vertical walls for the zero-padding, we use a {\it mesa-shaped} profile with
hangs smoothed via a cosine function. As usually done in the literature, we
first subtract the average of the map, then apply the zero-padding and raise
the result by the same amount we subtracted at the beginning. To comply with
the condition of equal treatment of the real and synthetic observations, we
apply the zero-padding to both kinds of datasets. The spectra obtained
through this method have less total energy, since we have 
flattened the velocity maps at their rim. To correct for this, we calculate
for each map the total energy before and after zero padding, and multiply the
power spectra of the zero-padded map with the corresponding factor. This is
akin to (and probably more accurate than) other $k$-independent corrections
used in the literature \citep[see, e.g.,][]{1987A&A...177..265V, rieutord2010}.

In the paper, in fact, averages over each time series (either observed or
simulated) are given instead of values for individual maps. To that end, we
first compute the power spectrum for each map, then calculate, for each $k$,
the average over the whole time series, $\langle P(k) \rangle$.  Finally, we
calculate the corresponding velocity spectrum through $\sqrt{k \langle P(k)
  \rangle}$. This averaging allows us to reduce the intermittence of individual
spectra by having a large statistical ensemble \citep[for a discussion, see
  also][]{Stein_etal_2006}. The kinetic energy spectrum for the statistical
ensemble can then be immediately obtained by averaging
Equation~\ref{eq:kinetic_energy_spectrum}.

\section{Fourier spectra for the vertical velocity}\label{sec:vertical}

\begin{figure*} [!ht]
\begin{center}
\includegraphics[width=1.0\textwidth]{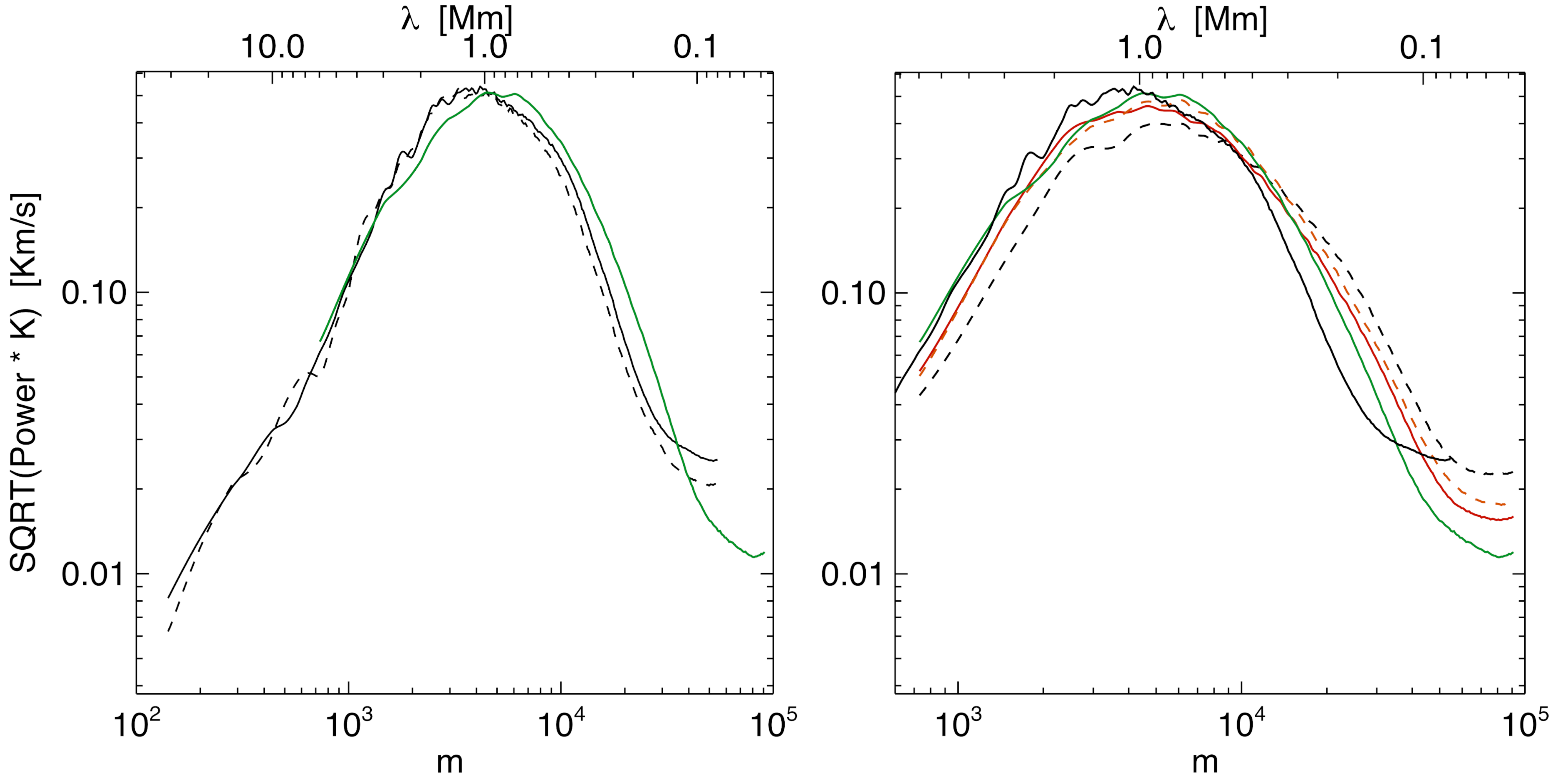}
\end{center}
\caption{Velocity spectra $V(k)$ of the Doppler velocity computed from the
  shift of the Stokes-I signal of the Fe I 5250.2 \AA\ line.
In abscissas we use the spherical harmonic wavenumber ($m$, lower horizontal
axis) or the corresponding wavelength $\lambda$ (upper horizontal axis). 
Left panel: curves corresponding to the first IMaX time series
(black dashed line); second IMaX time series (black solid line) and the HD
simulation run (green solid line). Right: curves corresponding to the
simulated HD/MHD time series  (green solid: HD; red solid: MHD 50G;
orange dashed: MHD 100G; black dashed: MHD 200G). The curve 
corresponding to the second IMaX time series
(black solid line) is added for comparison} \label{fig2}
\end{figure*}

\begin{figure*}[!ht]
\begin{center}
\includegraphics[width=1.0\textwidth]{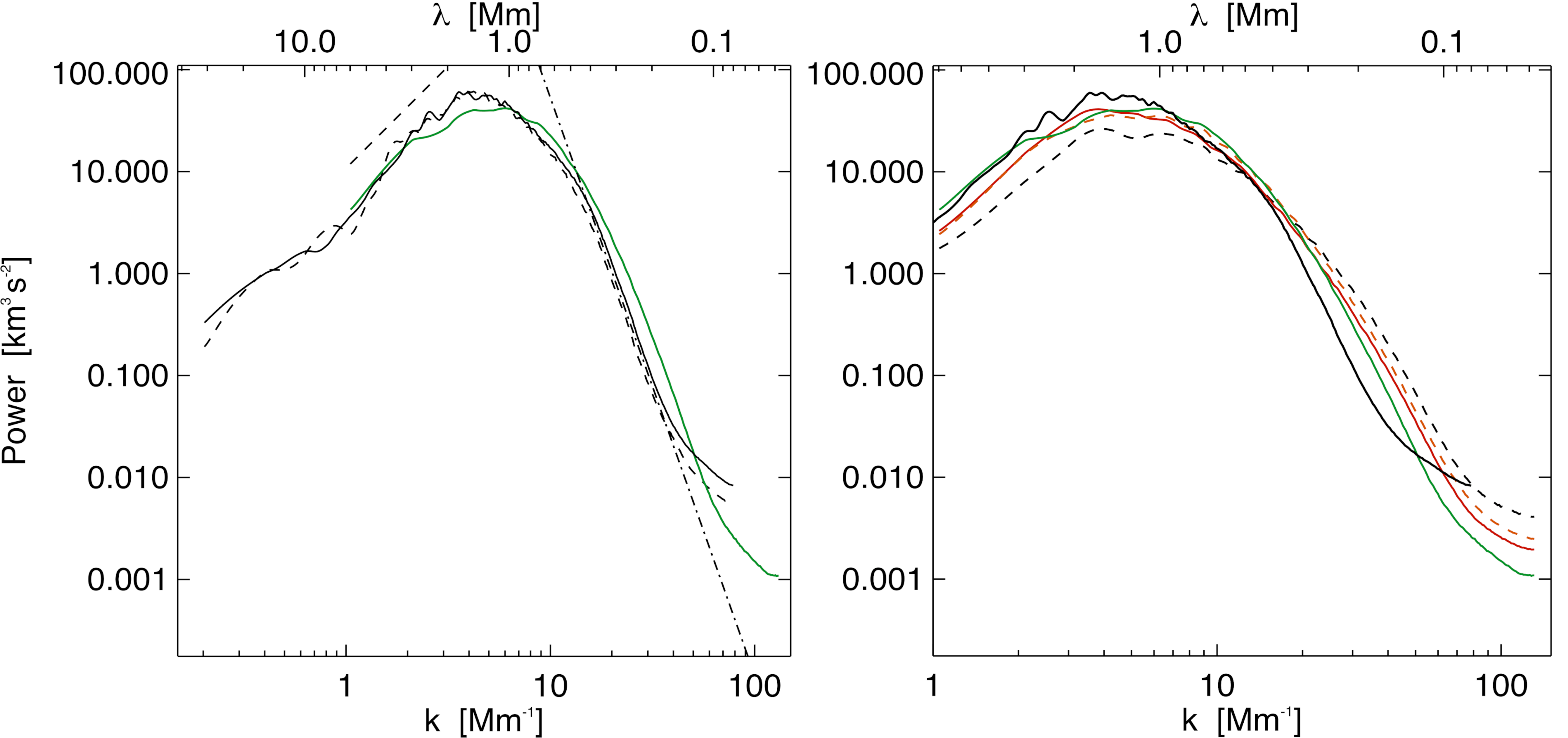}
\end{center}
\caption{Power spectra $P(k)$ of the Doppler velocity as a function of
  $k$. The curves here are for the same datasets as in Figure~\ref{fig2} and
  with the same color and linestyle criteria.  Left: spectra for the
  observational data and for the HD run. Right: the four simulated runs and
  the corresponding portion of the the second IMaX time series (black solid
  line). The straight lines in the left panel show power laws $\propto k^{2}$
  (dashed line), $\propto k^{-17/3}$ (dash-dotted) and $\propto k^{-29/3}$
  (solid).}
    \label{fig3}
\end{figure*}

\subsection{The velocity spectra $V(k)$}\label{sec:vel_spectra} 

The velocity spectra $V(k)$ for the vertical velocity in the two observed
IMaX time series are plotted in Figure~\ref{fig2} (left panel) as a function
of spherical harmonic wavenumber $m$ (lower horizontal axis) or wavelength
$\lambda$ (upper horizontal axis). The black dashed line corresponds to the
first time series, and the solid line is for the second
series. Additionally, we plot as a green solid line the spectrum
corresponding to the vertical velocity in the synthetic observation obtained
from the purely hydrodynamic (HD, i.e., $B=0$) numerical simulation.  In the
right panel, the spectra from the four numerical series are compared using the
following color coding: $\langle B_z \rangle=50$ G (red, solid), $\langle B_z
\rangle=100$ G (orange, dashed), $\langle B_z \rangle=200$ G (black, dashed),
with, again, the HD case in green.  In addition, the corresponding portion of
the second IMaX time series (black solid line) is plotted for comparison.

All the simulated and observed curves are in good general agreement for
scales ranging from $200$ km to $6$ Mm, i.e., up to the horizontal extent of
the simulation boxes. In fact, looking at the left-hand panel of the figure
(observations vs HD numerical simulations), the agreement can be said to be
excellent, considering that the two data sets have completely independent
origins (observations and simulations, respectively). There is a small
apparent shift in wavenumber between the simulated and the observational
curves; a shift of this kind was already obtained by \citet{Georgobiani2007}
(see their Figure~5) for the spatial scales allowed by the MDI data they used
in their work.  However, the total kinetic energy integral over all $k$
(~$\int V^2(k)\, d\,\hbox{ln}\,k$~) in the curves in Figure~\ref{fig2} (left
panel) is quite similar: the integral for the simulated HD series is between
the value of the integral for the two IMAX series, and the mutual deviation
between the latter is about $6\%$; that deviation is at any rate within the
observational and numerical errors of the curves.

\subsubsection{The granular range and larger spatial scales.}
All velocity spectra show a broad maximum in the granulation range, which
seems to be the only prominent scale in the figure. For the spectrum of the
observed data, for instance, the maximum has a value near $0.5$ km/s and is
located between $\lambda = 1$ and $1.2$ Mm.  For lengthscales larger than the
granular one (smaller $m$), the spectra roughly follow 
a power-law as a function of $m$ (see Sect.~\ref{sec:power_spectra}). None of
the curves shows any prominent mesogranular bump; this is particularly
interesting for the observational data, in spite of their large FOV. This
fits well with the results of \cite{yelles_etal_2011}. As for supergranular
features, their absence in the observational spectra may be caused by the
small horizontal extent of the sample: IMaX has a FOV of $32 \times 32$
Mm$^2$, so the observations probably do not contain enough supergranules.
The observations generally show higher power in the $1$ - $3$ Mm range than
the simulations (see the right panel of the figure), which may again be due
to the limited horizontal size of the simulation box.

\subsubsection{The subgranular range.}  
In the subgranular scale range, the observed velocity spectra (black curves
in the left panel) also decrease following an approximate power law
(Sect.~\ref{sec:power_spectra}) down to about $200$ km, where the slope
starts to become less negative.  This change is probably due to the proximity
of the effective spatial resolution limit ($\sim 120$ km) and also because of
numerical noise (arising, e.g., from the Doppler shift determination), which
is particularly clear for scales $\lambda < 120$ km.  A similar pattern is
followed by the simulation curves (right panel): in that case, however, there
is a clear inflection point located between $\lambda \sim 120$ and $140$ km.

\begin{figure*}[!ht]
\begin{center}
\includegraphics[width=1.0\textwidth]{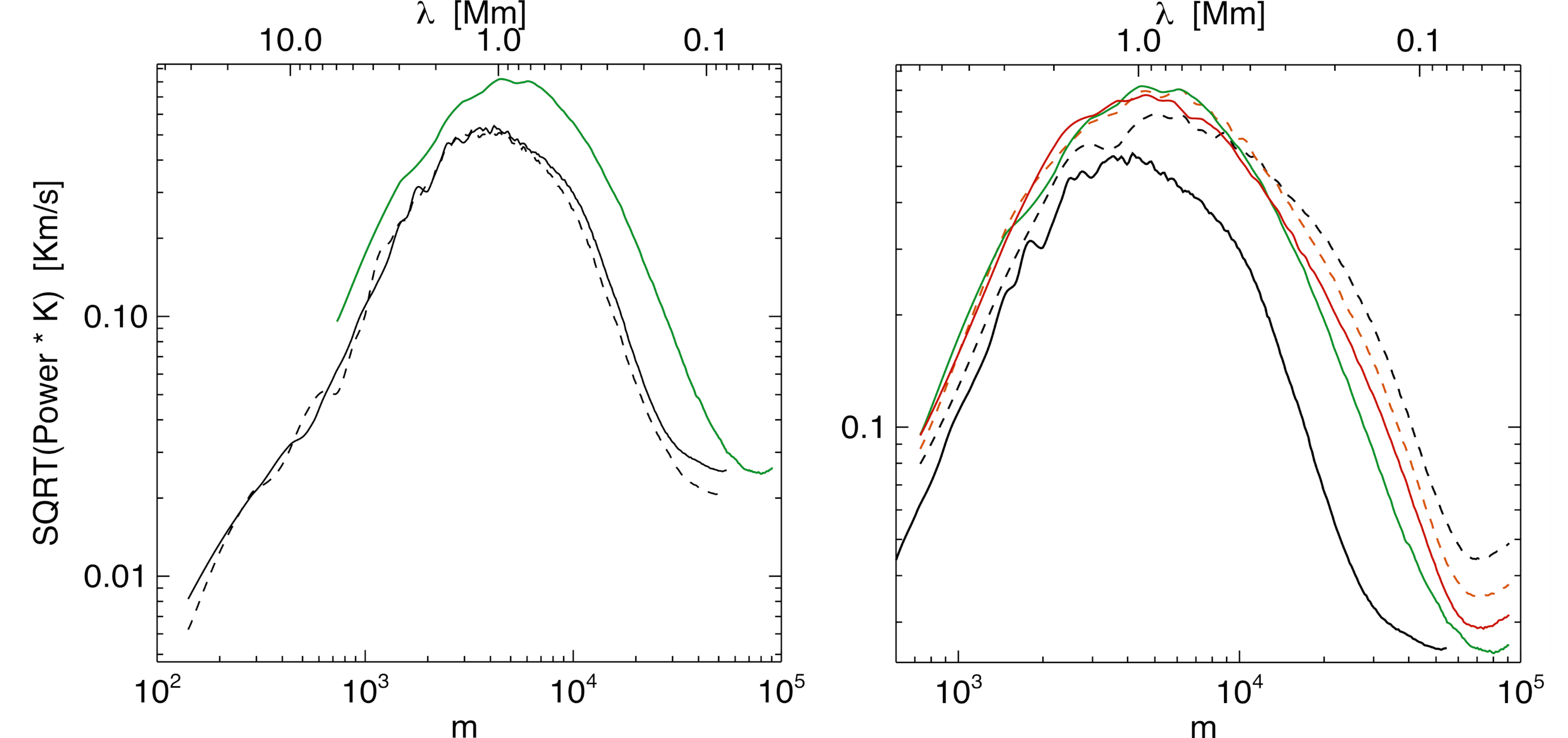}
\end{center}
\caption{Similar to Figure~\ref{fig2} with the simulated data not being
  convolved with any point spread function or spectral response
  function.} \label{fig4} 
\end{figure*}

\subsubsection{Comparison of the different magnetic
  runs}\label{sec:magnetic_runs}  

The right panel shows that the differences between the HD, $50$ G and $100$G
simulation cases are not large, with the $200$ G case showing the largest
deviation.  One can discern a clear trend in those curves: the curves have
larger power in the subgranular range the larger the average vertical field
$<B_z>$; for scales larger than about $400$ km, instead, the situation is
reversed. The integral over all $k$ in the figure reveals that the total
kinetic energy diminishes with increasing $<B_z>$: 
the normalized energy integral 
goes from $1$ (HD) to $0.87$ ($50$ G), $0.92$ ($100$ G) and $0.68$ ($200$ G),
which indicates that there is less total kinetic energy in the height range
of formation of the Fe I 5250.2 \AA\ line the larger the vertical magnetic flux.

\subsection{The power spectra $P(k)$}\label{sec:power_spectra} 

Figure~\ref{fig3} shows the power spectra of the Doppler
  velocity, $P(k)$, used in the calculation of
$V(k)$ for Figure~\ref{fig2}.  The curves of course have a similar
morphology to those of Figure~\ref{fig2} and their maximum
  occurs at a slightly larger horizontal scale ($\approx$ 1.5 Mm), as follows
  from the $\sqrt{k}$ multiplication when going from $P(k)$ to $V(k)$ (see
  Eq~\ref{eq:velocityspectrum}).  The dashed straight line in the left panel
  is a $k^{2}$ power law for comparison with the spectra on spatial scales
  above the granular one.  This value has also been obtained previously in
  the literature \citep[see, e.g.,][]{rieutord2010}. The dash-dotted straight
  line corresponds to a power law, $k^{-17/3}$, that fits the observed
  spectra for scales below the granulation down to $\approx$ 200 km
  (about where the observed spectra change their slope). \citet{rieutord2010}
  have found a similar value when using only two spectral wavelengths to
  calculate the Doppler velocity for this Fe I line but for scales larger
  than $\approx$ $400$ km. The tendency to find values for the slope more
  negative than $-5/3$ has also been reported by, e.g., \citet{Katsukawa2012}
  using Hinode/SP data and by \citet{Nordlund1997} using 3D MHD
  simulations. In fact the value $-5/3$ for the slope applies, strictly
  speaking, to fully isotropic turbulence \citep{Kolmogorov1941, Frisch1995,
    2008tufl.book.....L}, a condition that solar convection approaches to
  some extent only when going to subphotospheric levels
  \citep{Kitiashvili_etal_2012}.

\subsection{Fourier spectra without instrumental
  degradation}\label{sec:without_instrumental_degradation} The quality of the
match between observational and simulation curves apparent in
Figures~\ref{fig2} and \ref{fig3} would be clearly affected if we had not
taken care of the residual instrumental degradation. In Figure~\ref{fig4} we
show again the velocity spectra $V(k)$, but now the simulation results have
not been subjected to any degradation. There is a clear offset between the
two sets of curves. The kinetic energy of the HD case, for instance, is $2.5$
times larger than for the observations. Also: the maximum of the HD spectrum
has now a value near $0.8$ km/s. Given the excellent match of the two sets
when subjected to the same instrumental degradation, one can tentatively
conclude that the actual velocity spectra in the Sun are possibly not far
from the simulation ones in Figure~\ref{fig4} down to sizes of at least about
$100$ km and for $< B_z >$ in the range expected in the Sun, i.e., $< B_z
>\,\lesssim 100$ G.

\subsection{Comparison with previous results based on
    Hinode data}\label{sec:hinode} 

\begin{figure}[!ht]
\begin{center}
\includegraphics[width=0.5\textwidth]{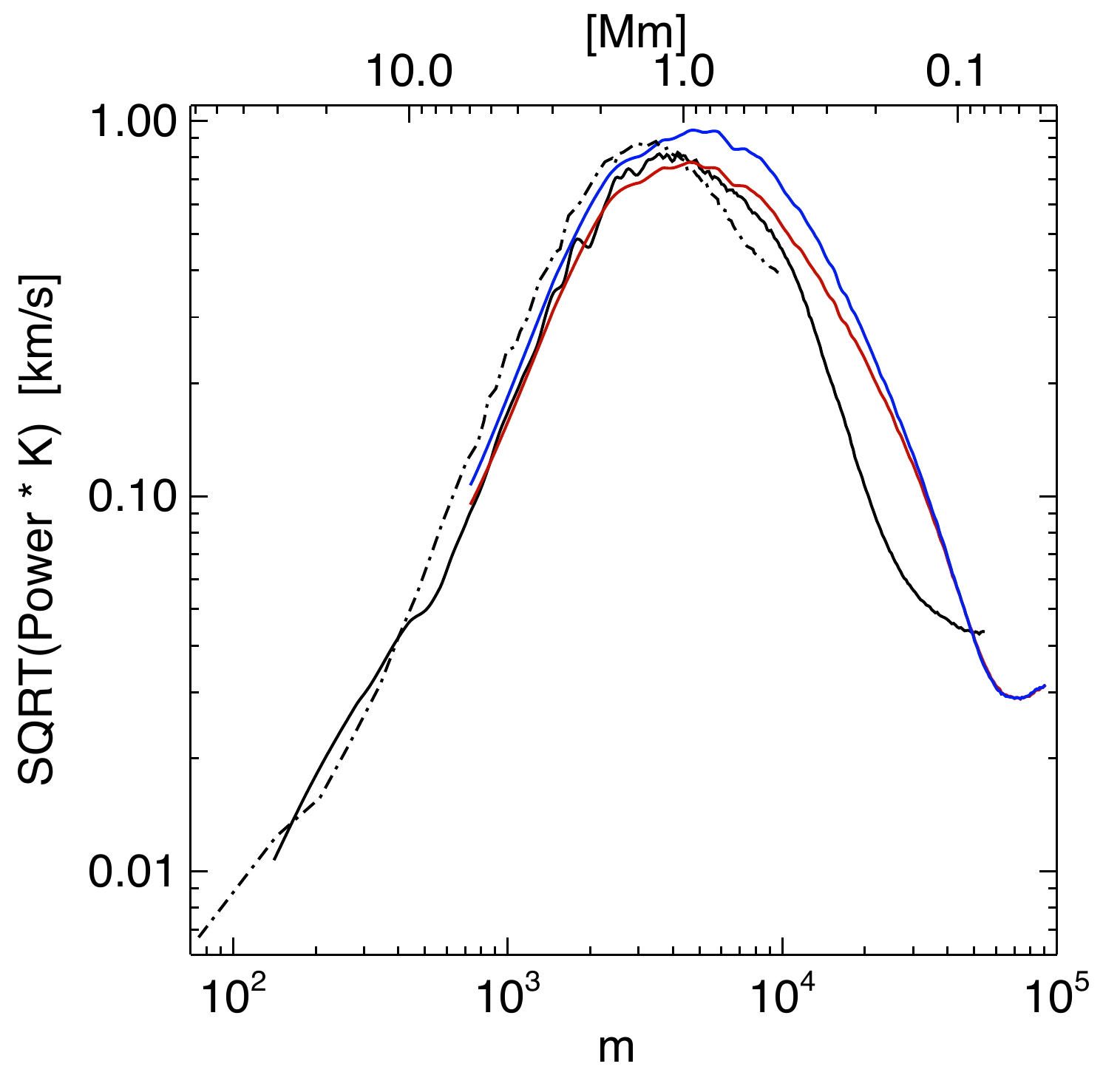}
\end{center}
\caption{Comparison with the Hinode-based Fourier spectrum of the Doppler
  velocity by \citet{rieutord2010}.  Solid black: spectrum from our second
  IMaX time series deconvolved with the $\psfprime$ function introduced in
  Section~\ref{sec:IMaXconditions}.  Dash-dotted: velocity spectrum computed
  from a power spectrum in Figure 6 of \citet{rieutord2010}, for a time
  series obtained from Hinode/NFI for the Fe I 5576.09 \AA\ line.  Red and
  blue curves: velocity spectra from our 50G MHD simulation run with
  spectral synthesis carried out for the Fe I 5250.2 \AA\ line and for the Fe
  I 5576.09 \AA\ line, respectively. To obtain those curves, we have not
  applied any instrumental degradation ($\psfprime$ and MSRF)
  to the simulated data.} \label{fig5} \end{figure}

For the particular case of the observed vertical velocities, Fourier spectra
have already been presented in the literature based on Hinode data
\citep[][]{rieutord2010} and it can be interesting to check for similarities
and differences with our results of Figures~\ref{fig2} and \ref{fig3}.  For the
comparison, we have taken the spectrum those authors present in the left
panel of their Figure~6, which corresponds to the kinetic energy associated
with the vertical component of the velocity, multiplied it by a factor $2$,
and obtained the corresponding velocity spectrum using our
Eq.~\ref{eq:velocityspectrum}. The result is plotted in Figure~\ref{fig5} as a
dash-dotted black curve. The curve of \citet{rieutord2010} had been treated
by those authors to remove instrumental effects via
deconvolution using a Point Spread Function that follows
  the prescription of \citet{Danilovic_etal_2008} for treating observations
  from SOT/Hinode.  For the comparison, we include in Figure~\ref{fig5} (black
  solid curve) the velocity spectrum corresponding to the second IMaX time
  series of Figure~\ref{fig2} but now deconvolved with the  $\psfprime$ function
  introduced in Section~\ref{sec:IMaXconditions}.
\footnote{Since we do not have available the Stokes-I
  images observed by IMaX in all wavelengths, we 
deconvolve  directly the velocity maps instead of
  the intensity maps.}  The comparison will be relevant if, as we expect, the
degradation effect on the IMaX data represented by $\psfprime$ is comparable
to the defocusing effect considered by \citet{Danilovic_etal_2008}. 
Furthermore, the Hinode data used by \citet{rieutord2010} were based on
Dopplergrams that used the Fe I 5576.09 \AA\ line, instead of (like IMaX) the
Fe I 5250.2 \AA\ one. Then, to complete the comparison we have computed the
velocity spectra from the simulated data ($50$-G MHD case) for those two Fe
lines and plotted them as red and blue curves (for Fe I 5250.2 \AA\ and Fe I
5576.09 \AA, respectively), of course without applying any instrumental
degradation to them.

Various features in this figure are worth commenting.  In broad terms, the
Hinode-based spectrum (dash-dotted curve) is quite similar to the other
spectra in the figure, especially for spatial scales $\lambda \gtrsim 1$
Mm. Comparing the simulation curves (blue and red lines) among themselves, we
see that the major difference between them is their amplitude near the
maximum. This may be associated with the different height of formation of the
FeI 5576.09 \AA\ and FeI 5250.2 \AA\ lines. Turning to the two curves in the
figure that use the FeI 5576.09 \AA\ spectral line (the Hinode dash-dotted
one, and the blue curve for our simulated data) there seems to be a
wavenumber shift between them. We assume that this feature may appear because
the effects of the lack of spatial resolution in the Hinode data start to
become apparent for $\lambda$ much larger than in the case of the simulated
data.  Another possibility may be the different characteristics (e.g.,
average magnetic field) of the region of the Hinode data compared to those of
the simulation. Comparing now the curves based on Hinode and IMaX
observations (dash-dotted and solid black), we see how the latter provides
useful information down to much smaller scales than the former: using, as a
reference, the inflection points in the subgranular range, we note that they
are located roughly around $700$ km (Hinode) and $300$ km (IMaX).  Turning to
scales above the granular one, the two curves show similar slopes, not far
from $\propto k^2$, and neither shows any prominent feature in that scale
range \citep[for the Hinode data, see discussion in][]{rieutord2010}.  A
final remark concerning the comparison in Figure~\ref{fig5}: the p-mode
filtering applied by \citet{rieutord2010} used a cut-off at $6$ km/s instead
of at $4$ km/s like in the present paper (as explained in
Sect.~\ref{sec:IMaXconditions}). We have checked that using either one
virtually makes no difference to the IMaX-based velocity spectrum (black
solid curve) and to the MHD-based curves (red and blue curves).

\begin{figure*}[!ht]
\begin{center}
\includegraphics[width=1.0\textwidth]{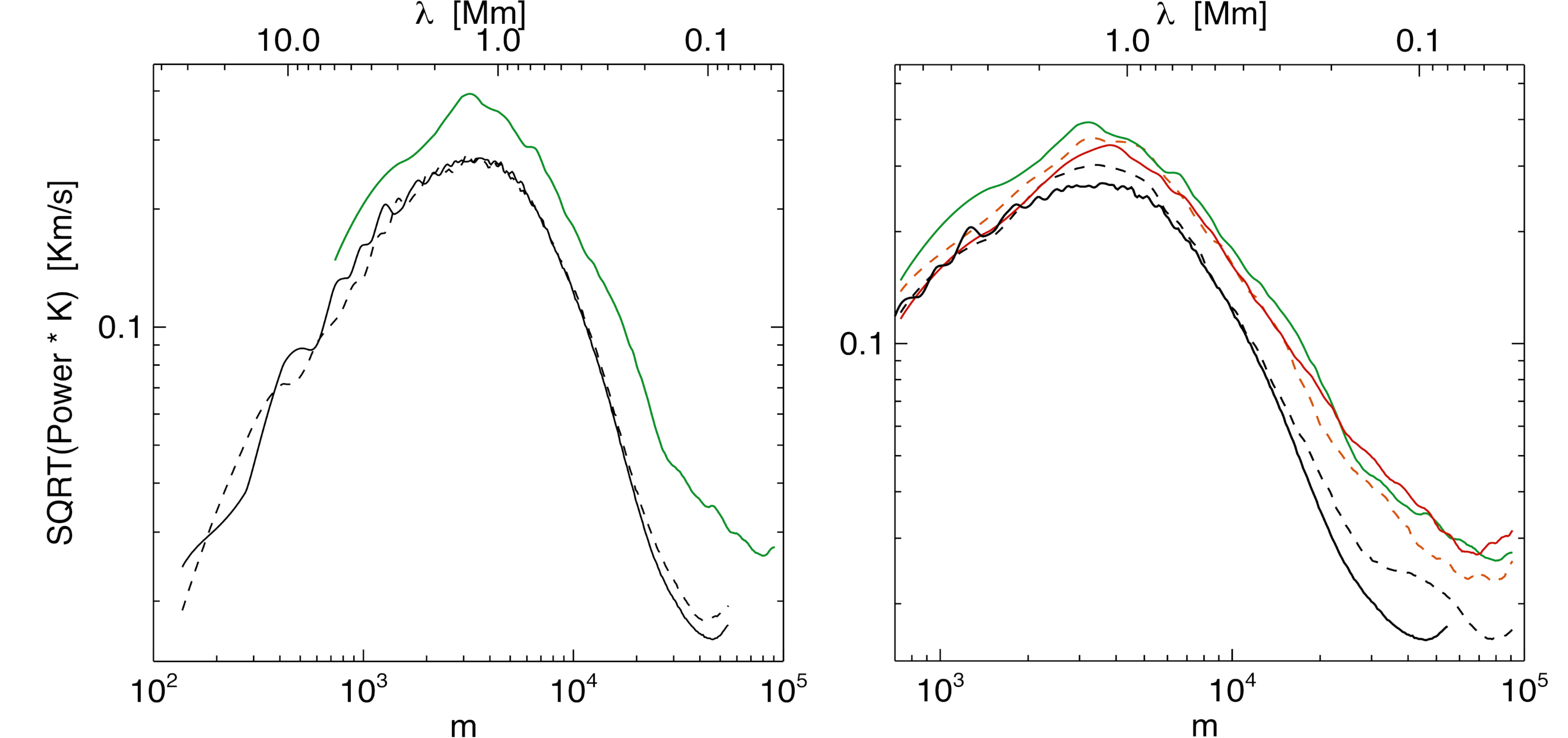}
\end{center}
\caption{Velocity spectra for the horizontal velocity computed through Local
  Correlation Tracking applied to continuum images near the Fe I 5250.2
  \AA\ line from the IMaX observations and from synthetic observations
  obtained from the simulation runs. Left panel: 
  first (black dashed) and second (black solid) IMaX time series and the HD
  simulation run (green solid). 
  Right panel: spectra from the four simulation runs (green: HD run; 
  red: MHD 50G run; orange dashed-line: MHD 100G run; black dashed-line: MHD 200G
  run). The corresponding portion of 
the second IMaX time series
(black solid line) is added for comparison.
} \label{fig6}
\end{figure*}

\section{Fourier spectra for the horizontal velocity}\label{sec:horizontal}

The horizontal velocity field for both the observed and simulated datasets
was determined through Local Correlation Tracking using 2D continuum maps for
a wavelength near the FeI 5250.2 \AA\ line
(Sect.~\ref{sec:IMaXconditions}). As discussed in the previous section, for a
proper comparison between observations and simulations, it is important that
we use the same kind of input data for the Fourier analysis: hence the need
to use synthetic observations from the simulated data (including instrumental
degradation), and to use the LCT technique the same way as with the IMaX
data. The extra advantage of having values for the actual velocity field in
all grid points across the simulation box will be exploited in the next
section (Sect.~\ref{sec:velocities_at_grid_points}). The LCT procedure has
often been applied in the literature to time-averaged data. The time
averaging is used to smooth the transition between consecutive snapshots and
reduce the noise. The averaging does not greatly affect the resulting
large-scale velocity field and power spectra (e.g., supergranulation), but
can considerably change the spectra in the small-scale (sub-granular) range
\citep[see examples and discussion in][]{rieutord2000, rieutord2001,
  rieutord2010, Stein_etal_2006, Kitiashvili_etal_2012}. The IMaX
observations have a high signal-to-noise ratio and a high spatial and
temporal resolution and suffer very little from atmospheric distortion. These
conditions make it possible to use LCT to determine instantaneous velocity
maps with no temporal averaging.  We can test the resulting velocity fields
concerning their behavior also in the small-scale range.  Like for the vertical
component, to obtain the power and velocity spectra for the horizontal
velocity field, we first compute the spectra of individual snapshots, and
then make an average of the spectra over the whole time series.

The velocity spectra $V(k)$ for the horizontal velocity are plotted in
Figure~\ref{fig6}. The color and line-type coding is the same as the one
employed in Figure~\ref{fig2}. The observed velocity spectra from the IMaX
time series (left panel, solid and dashed black curves) exhibit a maximum of
about $0.3$ km/s corresponding to horizontal scales around $1.3$-$1.4$ Mm.
The location of the maximum is quite similar to that obtained for the
vertical velocities.  In the right panel we see that all curves have a
similar shape but the observational ones exhibit less power as compared to
the simulated ones.  The green curve (HD case) is above all others, and
reaches a peak value of about $0.4$ km/s; the $200$G MHD simulation shows the
smallest values, with peak velocities of $0.3$ km/s. This confirms the
tendency obtained for the vertical velocity (Sect.~\ref{sec:vertical}),
namely, that the experiments with more magnetic flux exhibit less power, in
this case at heights where the continuum is formed. Here, however, this is
valid on all spatial scales (including the range below 300 km). In the
large-scale range, say above $2$ Mm, the observed and simulated spectra are
rather featureless: the corresponding power spectra $P(k)$ seem to follow a
power law as $k^1$. Again here, the only prominent scale in the whole range
is the granular one, but the caveat explained in Sect.~\ref{sec:vertical}
still applies: the limitation in the field of view does not allow us to study
properly the power on supergranulation scales. In the subgranular domain, the
observed spectra decrease smoothly with $m$ showing similar features as
described for the vertical velocity spectra in Sec~\ref{sec:vertical}: there
is an inflection point for  $\lambda \sim 300$ km (observations)
and $\sim 200$ km (simulations). Beyond this value the results are surely
contaminated by the remaining sources of noise.  In this case, the
corresponding power spectra $P(k)$ (not shown here) follow a power law as
$k^{-11/3}$ down to horizontal scales of about $200$ km.  This power law is
close to (but somewhat steeper than) that obtained from MHD simulations by
\citet{Kitiashvili_etal_2012}.  The observed and simulated spectra have
similar shapes, even if shifted with respect to each other.  However, in
Section~\ref{sec:velocities_at_grid_points}, we will see that the slope of
the power spectra of the LCT-determined horizontal velocity maps is strongly
influenced by the LCT procedure itself rather than reflecting the spectra of
the actual velocity field in the Sun.

\begin{figure}[!ht]
\begin{center}
\includegraphics[width=0.5\textwidth]{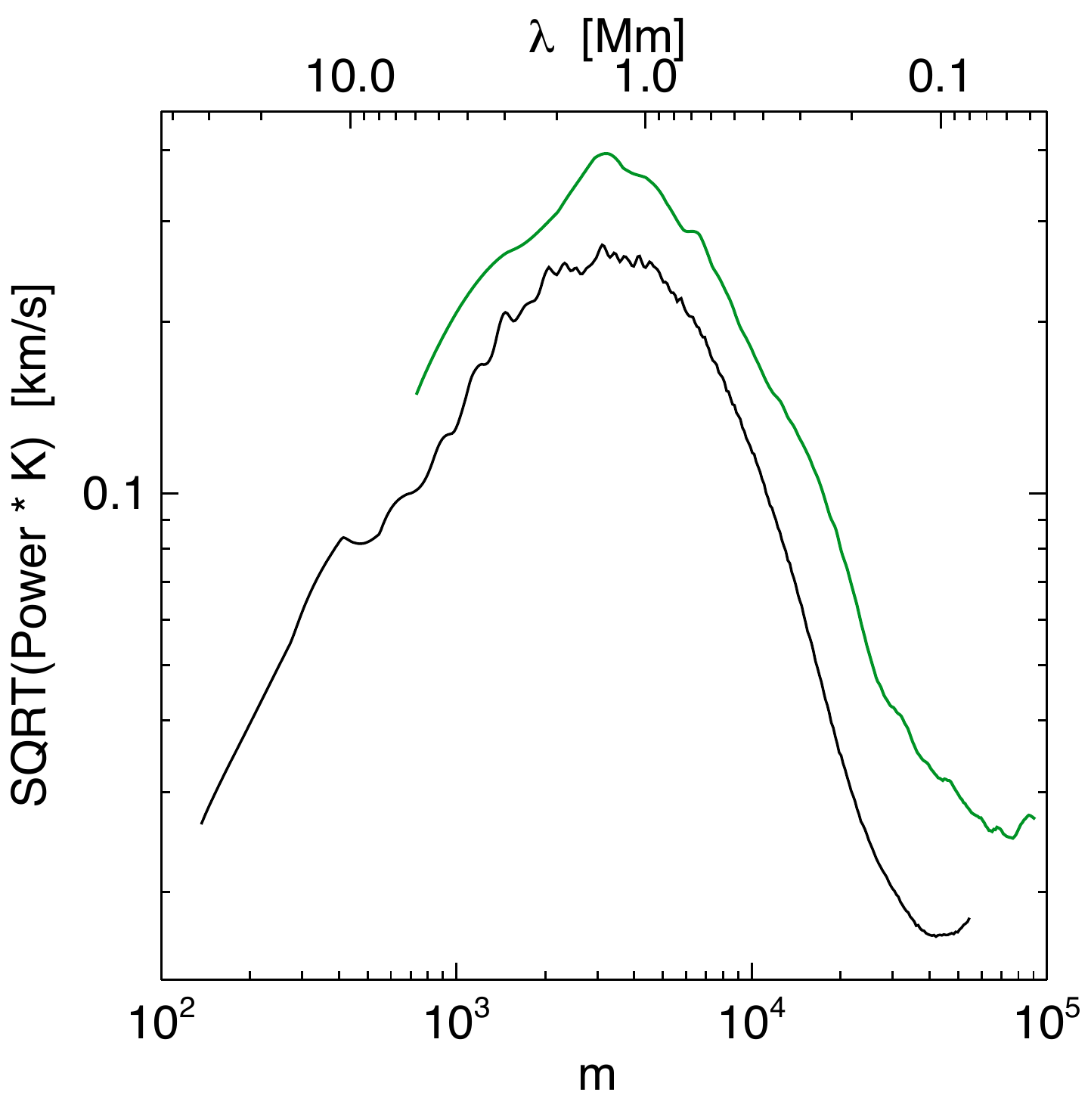}
\end{center}
\caption{Velocity spectra for the horizontal velocity computed from LCT
  eliminating the instrumental effects.  The black solid curve corresponds to
  the first IMaX time series deconvolved with the $\psfprime$ function
  introduced in Section~\ref{sec:IMaXconditions}.  The green line corresponds
  to the HD simulation run without application of any instrumental
  degradation.} \label{fig7}
\end{figure}

In Figure~\ref{fig7} we study the influence of the instrumental degradation
on the spectra (as we did for Figure~\ref{fig5}, but now without external
Hinode data).  The black solid curve corresponds to the first IMaX time
series deconvolved with the $\psfprime$ function introduced in
Section~\ref{sec:IMaXconditions}. It is very near the corresponding curve in
Figure~\ref{fig6}.  The green curve corresponds to the HD run; now, for
coherence in the figure, the LCT procedure has been applied to continuum
images that have not been convolved with $\psfprime$.  Also this curve is
quite near the corresponding one plotted in Figure~\ref{fig6}.  We conclude
that in this scale range the LCT procedure is not very sensitive to the
changes in the contrast of the continuum images: the convolution only
modifies the amplitude of the features in the continuum images without
obliterating them, so the tracking procedure yields similar velocity values.
In the sub-granular range, the two curves have similar slopes down to $\sim 200$
km.

\section{Comparing the velocities from the  synthetic observations with the
  velocity grid of the simulation cubes}\label{sec:velocities_at_grid_points}

In the previous sections, we have seen that the power and velocity spectra
from the simulations match those from the observations remarkably well in the
case of the vertical velocity. The match is less good for the horizontal
velocity. For the comparison, synthetic observations were obtained from the
simulations and then subjected to the residual instrumental degradation of
the IMaX maps.  Since we have the full 3D grid of velocity vectors in the
data cubes, it can be of interest to locate the positions in the cube where
the velocities best match the synthetic observations, both concerning Doppler
velocities as well as the LCT-determined horizontal velocities.  The way in
which the comparison is carried out has to be considered.  In the literature,
statistical studies directly based on the data grids in numerical simulations
have often been carried out \citep[e.g.,][]{Stein_Nordlund1998, rieutord2001,
  Stein_etal_2006, Georgobiani2007, matloch2010, Kitiashvili_etal_2012}, but
those calculations use velocity distributions on horizontal planes in the
box. For a comparison with observations (also with synthetic observations) it
seems more adequate to use velocity values taken on constant-$\tau_{500}$
surfaces: the standard velocity response functions (VRFs) of the $5250.2$
\AA\ and $5576.09$ \AA\ lines are expected to have roughly the same
dependence with $\tau_{500}$ in the different columns in the field of view
(FOV) \citep[e.g.,][]{gray2005}. In contrast, non-small vertical shifts of
the VRFs are expected between the different regions in the FOV when seen as a
function of the vertical coordinate, given the strong dependence of the
individual opacities on density. The velocities in a Dopplergram are
therefore expected to reflect the actual vertical velocity values on
$\tau_{500}$ isosurfaces more closely than those on horizontal planes. The
choice of the $\tau$-isosurfaces for the horizontal velocity determination is
less clear-cut, mostly because of the uncertainties in the LCT procedure, but
the choice of horizontal planes is equally questionable in that case, so we
have opted to use the same isosurfaces for both the vertical and horizontal
velocities.  In the following we first calculate the correlation between the
Doppler (or LCT-determined) velocities, on the one hand, and the simulation
data from individual $\tau_{500}=$const surfaces, on the other, so as to
determine which $\tau$ better corresponds to the observations. The
correlation is calculated for every individual snapshot in the series and
then averaged over all snapshots. In the case of the horizontal velocity, we
first convolve the velocity maps from constant $\tau_{500}$ surfaces with a
Gaussian with FWHM equal to the LCT tracking window size.  This allows us to
have both horizontal velocity maps at a similar spatial resolution.  As a
second item, we compare the Fourier spectra from either source. For the sake
of specificity, we use for this comparison the $100$-G MHD series.

In the case of the vertical velocity, it is found that the Doppler and MHD
velocities reach a maximum correlation on the order of $0.96$ near the
$\log(\tau_{500}) \approx -1$ surface. This indicates that the Doppler shift
measures the vertical velocity in that surface with a high accuracy; it also
shows that the Fe I 5250.2 \AA\ line is formed close to that optical
depth. The horizontal velocity field on const-$\tau$ surfaces and the
LCT-determined one reach a maximum correlation of close to $0.5$ at
$\log(\tau_{500}) \approx 0$. This $\tau_{500}^{}$ value is concomitant with
the fact that the LCT method applied here uses feature-tracking based on
continuum maps.  On the other hand, the lower value of the correlation occurs
because LCT as a method is less accurate in determining the
velocity than the Doppler procedure.  A precedent for this determination was
provided by \citet{rieutord2001} who computed the correlation between
LCT-determined horizontal velocity fields and velocity data {\it on
  horizontal surfaces} from a numerical simulation (they also used an
alternative to LCT, the CST method). They subjected their velocity data to
time averages and concluded that longer averaging time results in higher
correlations.  In our comparison, we use instantaneous velocity maps (with
$33$ sec cadence) and do not average the velocity data over time,
which explains our lower value for the maximum
  correlation (similar to what those authors obtain for $1000$-s
  averaging). We must use instantaneous determinations of the velocity since
  we want to study the power/velocity spectra almost down to
the diffraction limit of the telescope. 

\begin{figure}[!ht]
\begin{center}
\includegraphics[width=0.5\textwidth]{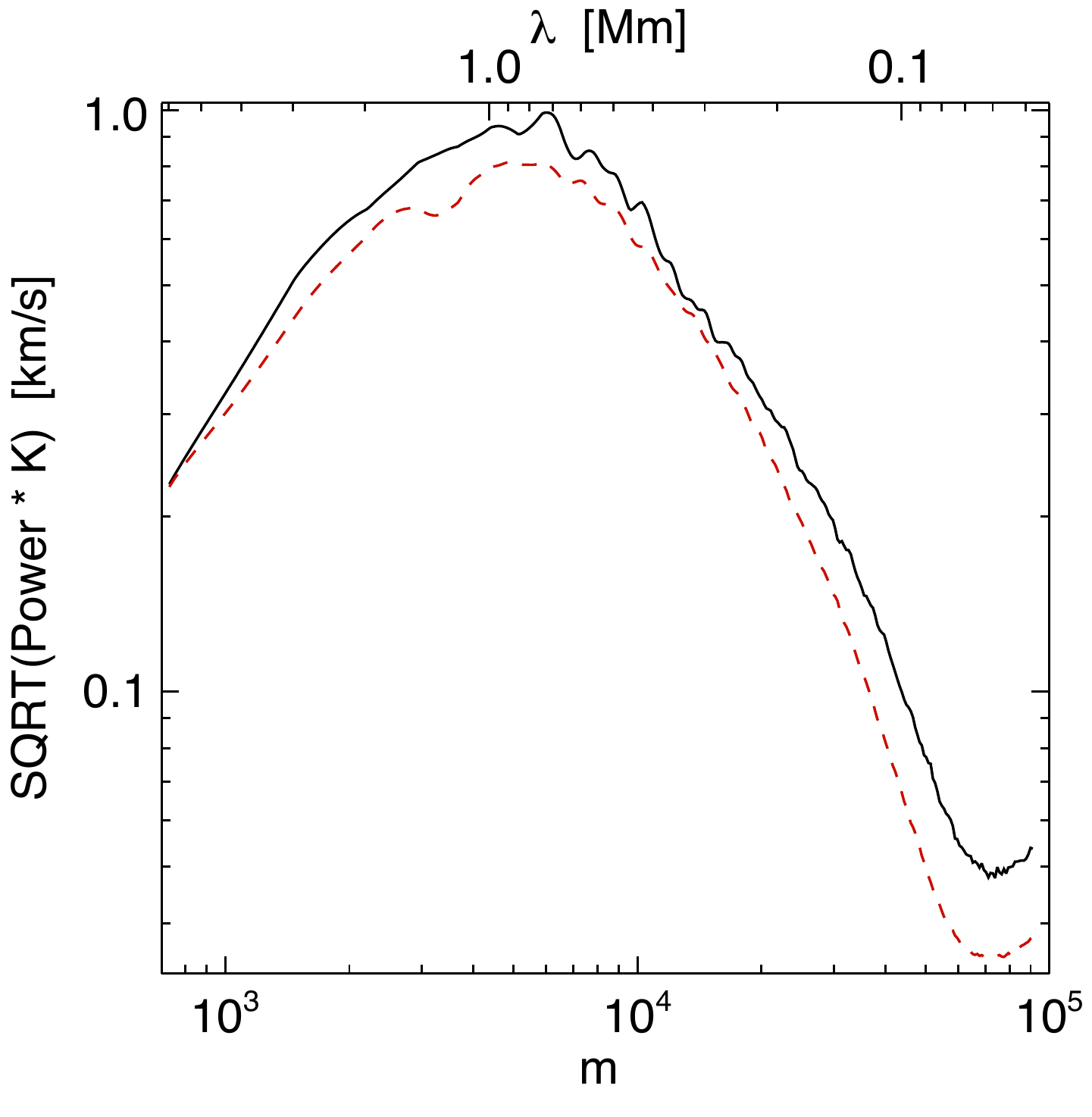}
\end{center}
\caption{Comparison between Fourier spectra from Doppler-determined
  velocities and from vertical velocities on constant-$\tau_{500}$
  surfaces, both obtained from the simulation data. Red: velocity spectra from
  the MHD 100G 
  data using Doppler shifts in the synthetic observation. 
  The data have not been convolved spatially with a PSF nor, spectrally, with
  an MSRF.  
  Black: velocity spectrum for the vertical component of the velocity taken
  directly from the MHD 100G data cube on the $\log(\tau_{500}) = -1$ surface. 
  No p-mode filtering has been applied to obtain this figure.} \label{fig8} 
\end{figure}

We now turn to the comparison of Fourier spectra for the data from the
simulations obtained using either synthetic observations or data from
constant-$\tau$ surfaces. Figure~\ref{fig8} shows velocity spectra for the
vertical velocity from the 100G MHD run. The dashed curve corresponds to the
synthetic observation, i.e., to the velocity measured through the Doppler
shift of the Fe I 5250.2 \AA\ line. The data used have not been subjected to
the instrumental degradation or spectral rebinning described in
Sect.~\ref{sec:IMaXconditions}, (i) -- (iii). The solid curve is for the data
taken directly from the MHD cube at the $\log(\tau_{500}) = -1$ surface,
where we found the maximum correlation earlier in this section.  The two
curves have a similar shape, with the one from the MHD cube exhibiting a
little more power on all scales. The ratio of the integrated kinetic energy
between the two datasets is $1.3$.  The agreement between the shape of the
two curves indicates that the Doppler measurement of the velocity is quite
reliably reproducing the velocity distribution over all spatial scales
included in the study, and thereby can be used to study the properties of the
photospheric turbulent flow in that range. For the spectra plotted in this
section, no p-mode filtering has been applied to facilitate the comparison
with the results from the data taken directly from the numerical box. Also,
since the simulation data are strictly periodic, we could have dispensed with
the zero-padding procedure explained in section
\ref{sec:fourier_definitions}, which is mandatory for the observational
data. However, the differences in the simulated spectra with or without it
are very small (in other words: the correction explained in that section is
quite accurate in this case), so we preferred to keep the zero padding and
apodizing also here to maintain coherence among the figures in the paper.

\begin{figure} [!ht]
\begin{center}
\includegraphics[width=0.5\textwidth]{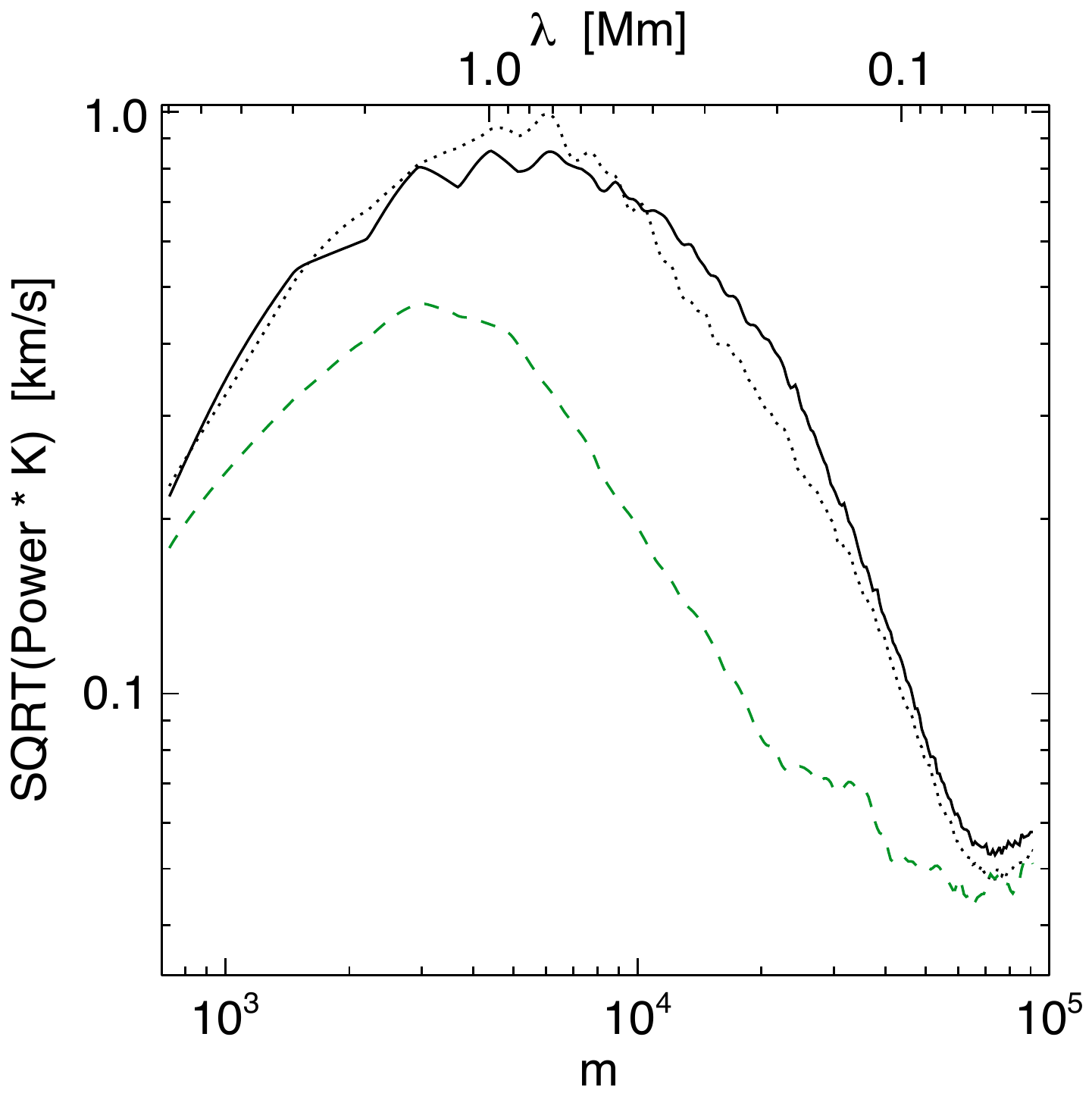}
\end{center}
\caption{Velocity spectra for the horizontal
  velocities.  Green dashed: spectrum corresponding to the horizontal
  velocity determined via LCT on continuum maps in the synthetic observations
  obtained from the MHD 100G simulation run. No instrumental degradation is 
  applied to the synthetic observations. Black solid: spectrum for the horizontal component of
  the velocity taken directly from the MHD 100 G data cube on the
  $\log(\tau_{500})=0$ surface. No p-mode filtering has been applied. 
  The vertical velocity spectra from the synthetic observations (taken from
  Figure~\ref{fig8}) is plotted for comparison (black dotted line)}  
\label{fig9}
\end{figure}

The corresponding spectra for the horizontal velocity are shown in
Figure~\ref{fig9}. The dashed curve is the velocity spectrum for the
LCT-determined horizontal velocity on the basis of the synthetic observation.
Like for the vertical velocity, no instrumental degradation has been applied to 
these synthetic observations. The solid curve corresponds to the spectra obtained from
the $\langle B_z \rangle = 100$G MHD cubes at $\log(\tau_{500}) = 0$, where the two data
sets have the highest correlation.  The latter curve shows much more power
than the LCT-based case: the ratio of the total energy between the two
horizontal velocity datasets is $5.4$.  This is in line with the concerns
expressed in the literature that the LCT method can importantly underestimate
the horizontal velocities \citep[e.g.,][]{1995ESASP.376b.219S,
  1995ESASP.376b.223S,   2006ApJ...638..553M}. 
Apart from the global shift downward, the LCT curve has a less negative slope
in the small-scale range, say for $\lambda \lesssim 300$ km. In fact the FWHM
of the window used here to determine the correlations in the LCT calculation
is $320$ km (Sect.~\ref{sec:IMaXconditions}), and this should be taken as a
lower bound below which the method cannot be expected to determine the
horizontal velocity field with any accuracy. Another related effect apparent
in the figure is that the dashed curve has a narrower maximum, and its peak
is at a larger horizontal scale (roughly $\lambda \sim 1.5$ Mm) than in the
case of the spectrum for the actual velocity field on the $\tau_{500}=1$
isosurface.  It is also of interest that the solid curves of
Figures~\ref{fig8} and \ref{fig9} have similar shapes (to facilitate the
comparison, we have reproduced the former as a dotted line in
Figure~\ref{fig9}). We see that the two curves are very near each other. The
coincidence, however, is of doubtful significance, since (a) the two curves
correspond to different (even if not far removed) heights in the photosphere
and (b) there need not be any vertical-horizontal isotropy in the convective
flows in those heights.
%

\section{Discussion}\label{sec:summary}

The availability in the past few years of very high resolution
spectropolarimetric observations of the solar photosphere through the
SUNRISE/IMaX mission has made it possible for us to explore the spectral
properties of the photospheric velocity fields down to scales of a tenth of
the average granular size, at the same time including large spatial scales,
up to the supergranular range (FOV of $33$ Mm).  Simultaneously, we have used
state-of-the-art, realistic models of solar magnetoconvection covering scales
of up to $6$ Mm to explore the level of agreement between the models and the
observations. In particular, we have compared the Fourier power spectra of
the velocity maps obtained from the IMaX time series with the ones obtained
from the MHD simulations both for the vertical component (via Doppler shift
determination) as well as for the horizontal component via Local Correlation
Tracking techniques. For the vertical velocity, when properly compared (i.e.,
when applying instrumental degradation to the simulations or, alternatively,
cleaning the observations from PSF effects), we have found a very good
agreement between observations and simulations (Figure~\ref{fig2} and
\ref{fig3}): the simulated spectra match the observed ones over a wide range
of scales, namely from $200$ km to $6$ Mm.  This lends additional credence to
the claim that present-day numerical simulations of solar convection provide
a faithful representation of the actual flows in the Sun down to the minimum
observable scales.  We have also found that the degradation of the
electromagnetic spectrum caused by the instrumental effects (PSF and MSRF)
can seriously modify the obtained kinetic energy of the flows (see
Figure~\ref{fig4}).  Still for the vertical velocity, we have compared the
Fourier spectra of the simulation data when using synthetic observations to
those obtained when using velocity values from $\tau_{500}$=const isosurfaces
(Sect.~\ref{sec:velocities_at_grid_points}). The excellent match between those
two spectra allows us to conclude tentatively that the non-degraded spectra
shown in Figure~\ref{fig4} may already be nearing the real solar Fourier
power spectrum for the vertical velocity in quiet Sun regions around the
formation height of the IMaX line (roughly $\log \tau_{500} \approx -1$, some
$200$ km above the nominal photosphere at $\log\tau_{500} = 0$) for spatial
sizes above the resolution limit of the data.  Additionally, from our spectra
one can also tentatively conclude that the typical granulation velocity in
those heights is around $0.8$ km/s.

Concerning the horizontal velocities, their determination in observations
through feature- or correlation-tracking in 2D maps is fraught with a
considerable uncertainty \citep[see, e.g.,][]{1995ESASP.376b.219S,
  1995ESASP.376b.223S, 2006ApJ...638..553M}.  With the tools developed for
this research, we have been able to compare the LCT velocity from the
synthetic observations in the MHD convection runs with the actual horizontal
component of the velocity on constant-$\tau_{500}$ surfaces in those same
simulations. We find that the maximum correlation between the LCT velocity
and the actual MHD horizontal velocity occurs near the solar surface
($\tau_{500} \approx 1$), which is reasonable since continuum maps are being
used for the LCT, and reaches comparatively low values of $\approx
0.5$. \citet{rieutord2001} showed that this kind of correlation (with
$v_{hor}$ calculated on horizontal planes in their case) can be increased by
carrying out a time-averaging in the LCT procedure with large time-windows
of, e.g., $1$ to $2$ hours, which reduces the small scale fluctuations.  In
our case we preferred to do without such time averaging not to dilute the
high-resolution properties of the IMaX data.

Another symptom of the inaccuracy of the LCT velocities can be seen in the
following: the velocity spectrum for the horizontal components of the
velocity on the $\tau_{500}=1$ surface (Figure~\ref{fig9}, black solid curve)
is well above the spectrum obtained from LCT-determined velocities on
synthetic observations from the same set of numerical simulations.  While the
former, in fact, has a shape similar to the spectrum for the vertical
velocities of Figure~\ref{fig8}, the latter shows a shift of its peak to
larger scales and a low value for the total energy.  This suggests that the
Fourier spectra from horizontal velocities determined through tracking
algorithms may not be fully representative of the solar reality in the
  range of length scales that can be studied with our numerical model.

A remarkable feature observed in Figures~\ref{fig2} -~\ref{fig6}, is the way
the power spectra are affected by the average magnetic flux in the simulation
boxes.  The cases with more magnetic flux have less power around the maximum
of the curves (Figure~\ref{fig4}, right panel). This behavior appears in the
spectra for both the vertical and horizontal velocities (Figures~\ref{fig2}
and~\ref{fig6}). An extension of this behavior toward the large-scale half of
the plots is particularly clear for the case with average magnetic flux
corresponding to a plage region ($200$ G case): it appears that the magnetic
field is partially inhibiting the convective flow on those scales \citep[see
  also the recent observational results by][]{Katsukawa2012}. In the
small-scale half of the plots for the vertical velocity, say for scales
$\lambda \lesssim 400$ km, the situation is the opposite: the spectra
(Figure~\ref{fig2}) show higher power the higher the average magnetic flux.
Together with the results concerning the total kinetic energy associated with
the velocity maps (Sect.~\ref{sec:magnetic_runs}), we tentatively conclude
that a higher average vertical magnetic flux leads to less kinetic energy and
to a shift in relative weight of the different scales toward the
small-$\lambda$ range. The behavior of the horizontal-velocity spectra in the
small-scale end seems to be at variance with these results, but, as concluded
in Sect.~\ref{sec:velocities_at_grid_points} (Figures~\ref{fig8} and
\ref{fig9}), the LCT-based results are not reliable enough in that part of
the spectrum.

An important outcome of any calculation of Fourier power spectra of surface
convection is the determination of the slope of the spectrum in the different
wavenumber regions of interest. In the figures of the present paper
approximate power laws can be discerned both on the large and small
wavenumber ranges on either side of the granulation maximum. From
Figure~\ref{fig3}, for instance, the power spectra for the vertical velocity
in the scale range from about $1$ Mm down to about $200$ km can be
approximated with a power law $k^{-17/3}$ for the IMaX series, and, less
accurately, for the HD simulation. The power spectra of the horizontal
velocity, on the other hand, also follow a non-classical power law of
approximately $k^{-11/3}$ on scales ranging from $\approx 1$ Mm down to $200$
km. Both exponents clearly deviate from the classical Kolmogorov $k^{-5/3}$
law valid for homogeneous and isotropic turbulence \citep{Kolmogorov1941,
  Frisch1995, 2008tufl.book.....L}. Additionally to the comments provided
when discussing Figure~\ref{fig3}, we note here that lack of spatial
resolution may be affecting the slope of the spectra: the figures in the
papers of \citet{Stein_Nordlund1998} (their Figure~30) and
\citet{Kitiashvili_etal_2012} (their Figure~2d) indicate that, for a grid
spacing as we have in our paper ($24$ km), the vertical velocity power
spectra show symptoms of nearing an inertial range with constant exponent at
subgranular wavenumbers.  The latter authors, in particular, obtain a
quasi-inertial range for the velocities on a horizontal cut at the surface
when using a grid spacing of $12.5$ km, albeit with a slope clearly steeper
than $-5/3$ (in fact, judging from their figure, probably between $-2$ and
$-3$). The same authors show how an extended inertial range with slope much
closer to $-5/3$ is obtained for the same resolution at much deeper levels
($3$ Mm below the solar surface), where isotropy should be more nearly
reached.

\section{Conclusions}\label{sec:conclusions}

 As measured using Fourier power spectra, realistic numerical
  simulations of surface magnetoconvection provide an accurate picture of the
  photospheric vertical velocity fields in the range from a few $100$ km to
  several Mm. This conclusion is based on a comparison of Dopplergrams
  obtained from real observations, on the one hand, and from synthetic
  observations calculated from the numerical data, on the other. Using as
  observational proxy for the horizontal velocities the Local Correlation
  Tracking method, the match is worse due to the limitations of that
  technique.

 Taking detailed care of spatial and spectral blurring caused by the
  observational equipment is necessary for a proper comparison between
  numerical and observational data: the instrumental degradation can
  seriously affect the Fourier power spectra of the LOS velocities (compare
  Figure~\ref{fig2} with Figure~\ref{fig4}), leading to a much worse match of
  synthetic and real observations. Apodization of the observational data
  prior to Fourier processing is advisable but less crucial.

 We have compared the vertical velocity proxy (Dopplergrams) in the
  synthetic observations with the actual velocity field on the individual
  $\tau_{500} =$ const isosurfaces in the numerical box. The maximum
  correlation ($0.96$) is found for the $\log\tau_{500}=-1$ surface, which is
  $\sim 200$ km above the continuum-forming layer ($\log\tau_{500}=0$). The
  corresponding Fourier spectra are very close to each other. The typical
  granulation velocity in those heights would be some $0.8-0.9$ km/s
  (Figure~\ref{fig8}).

 A similar comparison for the horizontal velocities leads to a smaller
 maximum correlation ($0.5$), which is reached at the $\tau_{500}=1$
 isosurface and to the spectrum for the actual velocity field on that
 isosurface being clearly above the spectrum for the LCT-determined velocity
 proxy (Figure~\ref{fig9}). Both problems can be ascribed to the limitations
 of the LCT technique particularly when dealing with small-scale features.

 Comparing cases with different magnetic flux, we tentatively conclude that a
 higher average vertical flux leads to less kinetic energy and to a shift in
 relative weight of the different scales toward the small-$\lambda$ range.

\begin{acknowledgements}
  We gratefully acknowledge financial support by the European Commission
  through the SOLAIRE Network (MTRN-CT-2006-035484) and by the Spanish
  Ministry of Research and Innovation through projects AYA2007-66502,
  CSD2007-00050, and AYA2011-24808. We acknowledge the computing time granted
  through the DEISA SolarAct and PRACE SunFlare projects and the
  corresponding use of the HLRS (Stuttgart, Germany) and FZJ-JSC JUROPA
  (J\"ulich, Germany) supercomputer installations, as well as the computer
  resources and assistance provided at the MareNostrum (BSC/CNS/RES, Spain)
  and LaPalma (IAC/RES, Spain) supercomputers. IMaX, an instrument on board the
  SUNRISE mission, was funded by the Spanish MICINN under projects
  ESP2006-13030-C06 and AYA2009-14105-C06. The authors are grateful to
  Dr.~H. Socas Navarro for providing the NICOLE spectral synthesis and
  inversion code and for his help in setting it up, to Dr A.~Nordlund for his
  encouragement to carry out this work and to Dr.~V. Mart\'\i nez Pillet for
  useful clarifications concerning the capabilities of the IMaX instrument.
  We are also grateful to Drs B.~Ruiz Cobo and D.~Fabbian for clarifications
  concerning the formation region of the spectral lines used in this paper
\end{acknowledgements}

\vskip 5mm


\end{document}